\documentstyle[12pt]{article}
\normalbaselines

\begin{document}
\begin{titlepage}

\vspace*{25mm}

\begin{center}
{\Large \bf Symmetries of the Self-Similar Potentials}%
\footnote{To appear in Comm.Theor.Phys. (Allahabad), vol. 2, 1993.}

\vspace{8mm}

{Vyacheslav Spiridonov%
\footnote{On leave of absence from the Institute for Nuclear Research,
Moscow, Russia}}

\vspace{2mm}
{\it Laboratoire de Physique Nucl\' eaire,
Universit\' e de Montr\' eal, \\
C.P. 6128, succ. A, Montr\' eal, Qu\' ebec, H3C 3J7, Canada}\\[5mm]
\end{center}

\vspace{6mm}

\begin{abstract}

An application of the particular type of nonlinear operator algebras
to spectral problems is outlined. These algebras are associated with
a set of one-dimensional self-similar potentials, arising due to the
$q$-periodic closure $f_{j+N}(x)=qf_j(qx),\; k_{j+N}=q^2 k_j$ of a
chain of coupled Riccati equations (dressing chain). Such closure
describes $q$-deformation of the finite-gap and related potentials.
The $N=1$ case corresponds to the $q$-oscillator spectrum generating algebra.
At $N=2$ one gets a $q$-conformal quantum mechanics,
and $N=3$ set of equations describes a deformation of the
Painlev\'e-IV transcendent.

\end{abstract}
\end{titlepage}

\newpage

Quantum algebras, or $q$-deformations of Lie algebras, attracted much
attention during the last years. They have appeared in conformal field
theories, spin-chain models, in the construction of link invariants,
and so on.
Another type of nonlinear algebras, known as $W$-algebras, also formed
the subject of recent intensive investigations. It is quite natural to try
to find applications of such objects within the context of
Sturm-Liouville type spectral problems.
This is not so straightforward and most of the attempts done
in this  direction cary phenomenological character -- the deformations of
spectra are introduced in an {\it ad hoc} manner, without proper definition
of arising operators.  As we shall show below, the problem can be treated
in a rigorous fashion on the basis of standard concepts of continuous space
physics.
Moreover, nonlinear algebras will appear very naturally as an inevitable
consequence of the analysis of exactly solvable potentials.

The consideration will be limited to the simplest one-dimensional
Schr\"odinger equation
\begin{equation}
-\psi^{\prime\prime}(x) + U(x) \psi(x) = \lambda \psi(x),
\label{a}
\end{equation}
describing a particle moving in some potential $U(x)$.
(The prime in (\ref{a}) and below always denotes derivative with respect to
real coordinate $x$, $x\in R$.)
The quantum mechanical spectral problems, associated to (\ref{a}),
traditionally provide good place
for probing new group-theoretical ideas. This is inspired by the general
belief that any kind of regularity in spectral data is generated by some
symmetry algebra of a Hamiltonian. The qualitative
understanding of peculiarities of a given system is reached when an operator
algebra governing the map of physical states onto each other is explicitly
constructed. The most advanced approach to building of such symmetry
transformations, which we are
going to exhibit, is connected with the technique developed by Darboux
long time ago \cite{r1}. Within quantum mechanics, it is known as
factorization method \cite{r2}. In the theory of
integrable nonlinear evolution equations, it was generalized and named as the
dressing method \cite{r3}. Later, Darboux transformations were identified as
supersymmetry transformations, mixing bosonic and fermionic degrees of freedom
in specific models \cite{r4}. Some further parastatistical
generalization of the latter interpretation was suggested in Ref.\cite{r5}.
Using this method, we shall describe a set of self-similar
one-dimensional potentials whose discrete spectra are composed from a number
of geometric series. Among
the corresponding nonlinear spectrum generating algebras we shall find
a $q$-deformed Heisenberg-Weyl algebra and a quantum conformal algebra
$su_q(1,1)$, where parameter $q$ will have the meaning of interaction
constant.

Quantization of the spectral parameter $\lambda$ emerges due to
particular boundary conditions imposed upon the wave functions $\psi(x)$.
Just for an illustration we exhibit one of the
possible (self-adjoint) conditions:
\begin{equation}
\psi^\prime(x_1)=h_1 \psi(x_1), \qquad \psi^\prime(x_2)=h_2 \psi(x_2),
\label{b}
\end{equation}
where $x_1$ and $x_2$ are two different points on the line, and $h_1,\, h_2$
are arbitrary real constants. Here we would like to mention the powerful
restriction on asymptotic growth of eigenvalues $\{\lambda_n\}$ for
the potentials without singularities in $[x_1, x_2]$. Namely,
$\{\lambda_n\}$ can not grow faster than $n^2$
at $n\to \infty$ \cite{r5a}. This observation
immediately restricts the region of applications of $q$-deformed
commutation relations which often formally produce exponentially growing
discrete spectra. Below we imply the condition of square normalizability
of wave functions, $\psi(x) \in L^2 [-\infty,\infty]$, for the discrete
energy states. For singular potentials one still will need
additional boundary conditions. The (quasi)periodic potentials
with zonal structure of spectra will also appear in the consideration.

Let us first present some basic principles of the factorization method.
The main object appearing after successive factorizations of the stationary
one-dimensional Schr\" odinger operator (\ref{a}) is the following chain of
coupled Riccati equations (the dressing chain):
\begin{equation}
f_j^\prime(x)+f_{j+1}^\prime(x)+f_j^2(x)-f_{j+1}^2(x)=
k_j\equiv \lambda_{j+1}- \lambda_j,\qquad j\in {\it Z}
\label{e1}
\end{equation}
where $k_j\, (\lambda_j)$ are some constants. The Hamiltonians $L_j$
associated to (\ref{e1}) have the form,
\begin{equation}
L_j=p^2+f^2_j(x)-f^\prime_j(x)+\lambda_j= a^+_ja^-_j +\lambda_j,
\label{e2}
\end{equation}
$$a_j^\pm=p\pm if_j(x),  \qquad p\equiv -id/dx.$$
Conditional isospectrality of $L_j$ and $L_{j+1}$ follows from the
intertwining  relations
$L_j a_j^+=a_j^+L_{j+1},$ $a_j^-L_j =L_{j+1}a_j^-.$
Any exactly solvable spectral problem with infinite number of discrete
levels can be represented in the form (\ref{e1}), (\ref{e2})
with $\lambda_j$ being the Hamiltonian eigenvalues:
$$L_j\psi_n^{(j)}=E_n^{(j)}\psi_n^{(j)}, \qquad
E_n^{(j)}=\lambda_{n+j},\quad n=0,1,2,\dots$$
\begin{equation}
\psi_n^{(j)}\propto a_j^+\psi_{n-1}^{(j+1)},\qquad
\psi_{n-1}^{(j+1)}\propto a_j^-\psi_n^{(j)},
\label{e3}
\end{equation}
$$\psi_{0}^{(j)}(x)=\exp{(-\int^x f_j(y)dy)}.$$
In this case $\lambda_{j+1}>\lambda_j$ and all wave functions
$\psi_n^{(j)}$ are square integrable, with $n$ nodes inside taken
coordinate region.
For Hamiltonians with finite number of bound states normalizability
of $\psi_0^{(j)}$ truncates at some $j$. A large list of potentials,
whose spectra are easily found by the factorization method, is given
in the first paper
of Ref.\cite{r2}. In general, however, parameters $\lambda_j$ do
not coincide with physical eigenvalues since for a given potential they
could be chosen arbitrarily.

In order to solve underdetermined system (\ref{e1}) one has to
impose some closure conditions. At this stage it is an art of a researcher
to find such an ansatz, which allows to generate infinite number of
$f_j$ and $k_j$ by simple recurrence relations from fewer entries.
Most of the old known examples
are generated by the choice $f_j(x)=a(x)j +b(x) +c(x)/j$, where $a,b,c$ are
some functions determined from setting powers of $j$ in (\ref{e1})
equal to zero.
Three years ago Shabat and Yamilov considered the problem from another
point of view \cite{r6}.
They suggested to treat (\ref{e1}) as some infinite-dimensional dynamical
system and to find potentials corresponding to finite-dimensional
truncations of the chain. In particular, it was shown that very simple
periodic closure conditions:
\begin{equation}
f_{j+N}(x)=f_j(x),\qquad \lambda_{j+N}=\lambda_j,
\label{e4}
\end{equation}
for $N$ odd lead to the potentials with zonal structure of spectra
which are described by (hyper)elliptic functions and are called the finite-gap
potentials (parameters $\lambda_j$ represent physical
eigenvalues when they coincide
with the boundaries of gaps). First non-trivial example appears
at $N=3$ and corresponds to the one-gap Lam\'e equation.

Just slight modification of (\ref{e4}):
\begin{equation}
f_{j+N}(x)=f_j(x),\qquad   k_{j+N}=k_j,
\label{e6}
\end{equation}
describes essentially more complicated potentials, although
one has only one additional free parameter $\omega=k_j+k_{j+1}+\dots
+k_{j+N-1}=\lambda_{j+N}-\lambda_j\neq 0$.
It is easy to see that
(\ref{e6}) at $N=1$ gives the harmonic oscillator problem,
whereas (\ref{e4}) is degenerate. The $N=2$ system coincides with the general
conformal quantum mechanical model \cite{r7a},
\begin{equation}
f_{1,2}(x)={1\over 2}(\pm\,{k_1-k_2\over k_1+k_2}{1\over x}+{k_1+k_2\over 2}x).
\label{e7}
\end{equation}
Already $N=3$ case leads to transcendental potentials \cite{r7}, namely,
$f_j(x)$ depend now on
solutions of the Painlev\'e-IV equation \cite{r1}:
\begin{equation}
f_1(x)={1\over 2}\omega x+f(x),\qquad
f_{2,3}(x)=-{1\over 2} f\mp \frac{1}{2f}(f^\prime+k_2),
\label{e8}
\end{equation}
\begin{equation}
f^{\prime\prime}={{f^\prime}^2\over 2f}+{3\over2}f^3+2\omega x f^2+
(\mbox{$\frac{1}{2}$}\omega^2 x^2+k_3-k_1)f - {k_2^2\over 2f}.
\label{e9}
\end{equation}
To the author's knowledge this is the first example, when Painlev\'e
transcendent appears in a quantum mechanical context.

Let us take two unconstrained
Hamiltonians $L_j$ and $L_{j+N}$ from the chain (\ref{e2})
and assume that both are self-adjoint.
The map from unnormalized wave functions of
$L_{j+N}$ to those of $L_j$ is performed by successive action
of the operators $a^+_l$,
$$
\psi^{(j)}_{n^\prime}(x)=K_j^+\psi_{n}^{(j+N)}(x),\qquad
K_j^+=a_j^+ a_{j+1}^+\dots a_{j+N-1}^+, \quad K_j^-\equiv (K_j^+)^{\dag}
$$
\begin{equation}
K_j^+ L_{j+N}=L_j K_j^+,\qquad L_{j+N}K_j^-=K_j^- L_j.
\label{e10}
\end{equation}
For a discrete spectrum the labelling of levels differs by some integer,
$n^\prime =n+m,\, |m|\leq N$.
Note that for superpotentials $f_j$ with strong enough singularities
some of the Hamiltonians $L_j$ are not
isospectral because the
operators $a^\pm_l$ are not well defined for a subset of physical states.

Now we restrict potential of $L_{j+N}$ to be equal to that of $L_j$
up to some simple transformation.
Then  $K^\pm_j$ map  eigenstates of $L_j$ onto each other,
$\psi_{n^\prime}^{(j)}\leftrightarrow \psi^{(j)}_n$ (there may be zero modes).
The simplest case is realized
when all states are mapped onto themselves, i.e. when $n^\prime=n$. This
is equivalent to $L_{j+N}=L_j$, or (\ref{e4}). From (\ref{e10}), the
operators $K^\pm_j$
are seen to be integrals of motion \cite{r6}.
For $n^\prime\neq n$ a spectrum generating situation arises.
Taking $\lambda_{j+N}\neq \lambda_j$ and substituting
$L_{j+N}=L_j+ \lambda_{j+N}-\lambda_j$ into (\ref{e10}) one gets the standard
ladder relations. Integrability of these periodic truncations of
the chain (\ref{e1}) is analyzed in Refs.\cite{r6,r7}. Any further
generalizations of the results mentioned above would involve
solvable potentials with infinite number of levels (gaps).
Let us show that the notion of $q$-deformation naturally emerges on this
track.

A peculiar potential was found in Ref.\cite{r8}
by the following self-similarity constraint:
\begin{equation}
f_j(x)=q^j f(q^j x),\qquad k_j= q^{2j}k, \qquad 0<q<1, \quad k>0,
\label{e11}
\end{equation}
which gives a solution of (\ref{e1}) provided $f(x)$ satisfies the equation
\begin{equation}
f^\prime(x)+qf^\prime(qx)+f^2(x)-q^2f^2(qx)=k.
\label{e12}
\end{equation}
Quantum algebraic content of this model was uncovered in Ref.\cite{r9},
where it was shown that $q$-deformed Heisenberg-Weyl algebra \cite{r10},
\begin{equation}
A^-A^+-q^2 A^+A^-=k,
\label{e13}
\end{equation}
is implemented by the choice
$$A^+=(p+if(x))T_q,\quad A^-=T_q^{-1}(p-if(x))=(A^+)^{\dag},$$
\begin{equation}
T_qf(x)=\sqrt{q}f(qx).
\label{e12a}
\end{equation}
Intertwining relations between $A^\pm$ and Hamiltonian
$H=A^+A^- -k/(1-q^2)$ easily generate the spectrum
\begin{equation}
H A^\pm =q^{\pm 2}A^\pm H, \quad \Rightarrow \quad E_n=-k q^{2n} /(1-q^2).
\label{e14}
\end{equation}

A deformation of supersymmetric quantum mechanics, inspired by this model,
was suggested in Ref.\cite{r11}. The main idea is very simple -- one
has to replace superpartner Hamiltonian by that obtained after
affine transformation (i.e. dilatation  and translation) and adjust kinetic
term
to the standard form. Degeneracies
of levels are removed and energy split is proportional to $1-q^2$,
where $q$ is the scaling parameter. Within this scheme, Eq.(\ref{e12})
is a condition of homegeneity of magnetic field alond third axis for
a spin-1/2 particle moving on the line.
This construction is easily generalized to the particular parasupersymmetric
model defined by unification of sequential members of the chain
(\ref{e2}) into diagonal $(N+1)\times (N+1)$ matrix. Acting on each
subhamiltonian by different affine transformation group elements and
rearranging kinetic terms one would get
multiparameter deformation of parasupersymmetric algebraic relations.
Following the consideration of Ref.\cite{r5}, one may impose various physical
restrictions on the matrix Hamiltonian and look for explicit
form of  potentials accepting those constraints. Analyzing such possibilities
the author have found
the following $q$-periodic closure of the chain (\ref{e1}),
\begin{equation}
f_{j+N}(x)=qf_j(qx), \qquad k_{j+N}=q^2 k_j.
\label{e15}
\end{equation}
It leads to a set of mixed finite-difference-differential equations
which describes $q$-deformation of the finite-gap and related
potentials discussed in Refs.\cite{r6,r7}.
The self-similar system of Ref.\cite{r8} is generated at $N=1$.
Because of the highly
transcendental character of self-similarity and connections with the
Painlev\'e equations
all corresponding potentials may be called as $q$-transcendental ones.

Let us find a symmetry algebra  behind (\ref{e15}).
First we rewrite $q$-periodicity at the Hamiltonian level:
\begin{equation}
L_{j+N}=q^2 T_q L_j T_q^{-1} +\sigma_j, \qquad
\sigma_j=\lambda_{j+N}-q^2\lambda_j,
\label{e16}
\end{equation}
where we normalize $\sigma_j>0$. Substituting this into (\ref{e10}) we get
\begin{equation}
L_j A_j^+-q^2A_j^+L_j = \sigma_j A_j^+, \qquad
A_j^-L_j  -q^2L_jA_j^- = \sigma_j  A_j^-,
\label{e17}
\end{equation}
$$A_j^+\equiv K_j^+ T_q, \qquad A_j^-=(A_j^+)^{\dag}.$$
Obtained formulae represent a first part of the quantum algebraic relations
determining structure of the system. Second part is fixed by a particular
$q$-commutator of $A_j^\pm$ following from the identities:
\begin{equation}
A^+_jA^-_j =\prod_{i=0}^{N-1} (L_j-\lambda_{j+i}),\qquad
A^-_j A^+_j=\prod_{i=0}^{N-1} (q^2 L_j+\sigma_j-\lambda_{j+i}).
\label{e18}
\end{equation}
As an example, we write one possible equality:
\begin{eqnarray}
A_j^+A^-_j -q^{2N}A^-_jA^+_j= {\cal P}_{N-1}(L_j),
\label{e19}
\end{eqnarray}
where ${\cal P}_{N-1}$ is a polynomial of the degree $N-1$.
Formulae (\ref{e17}), (\ref{e19}) define a particular class of
nonlinear algebras
which may be interpreted as a $q$-deformation of the polynomial algebra
of ordinary differential operators discussed in Ref.\cite{r7} (the latter
in turn may be considered as the simplest $W$-algebras).
Corresponding ladder relations determine the spectrum provided
$A_j^\pm$ respect boundary conditions of a problem.
The peculiarities of representations of general
nonlinear algebra $[J_0, J_\pm ] = \pm J_\pm,\; [J_+, J_-]=g(J_0)$, where
$J_{\pm,0}$ are some formal operators, were
discussed in Ref.\cite{roc}.

For simplicity, we restrict our consideration to the whole line spectral
problems with non-singular potentials, in which case
all $a^\pm_j$'s are well defined.
Then the index $j$ can always be chosen in such a way that $A_j^+$ will not
have zero modes (this will be assumed below).
Normalizability of physical states
is not spoiled by $A_j^\pm$-operators, which thus
raise and lower energy.  As a result, the equation
$A_j^- \psi(x)=0$ determines lowest energy state. Suppose that all $N$
independent solutions of this equation are normalizable, which corresponds to
the ordering  $\lambda_j<\lambda_{j+1}<\dots <\lambda_{j+N-1}$ and
normalizable  $\psi_0^{(l)}$'s in (\ref{e3}). Energies of all bound states
are easily found:
\begin{equation}
E_n^{(j)}={\sigma_j\over 1-q^2}+ \cases{
(\lambda_j-{\sigma_j\over 1-q^2})q^{2r}, & for $ n=Nr$\cr\noalign{\vskip2pt}
(\lambda_{j+1}-{\sigma_j\over 1-q^2})q^{2r},
&for $n=Nr+1$\cr\noalign{\vskip2pt}
\dots&\dots\cr\noalign{\vskip2pt}
(\lambda_{j+N-1}-{\sigma_j\over 1-q^2})q^{2r},& for $n=Nr+N-1$\cr}
\label{e20}
\end{equation}
By definition $E_n^{(j)}<E_{n+1}^{(j)}$, and since $q<1$, one
has $E^{(j)}_\infty=\sigma_j/(1-q^2), \; f_\infty(x)=0$,
i.e. the potentials are reflectionless,
with spectra comprising $N$ geometric series.
For the continuous spectrum the roles of $A^\pm_j$ are interchanged,
action of $A^-_j$ creates a series of states with exponentially
growing eigenvalues.
Note that at $q\to 0$ there will remain only first $N$
levels corresponding to zero value of $r$ in (\ref{e20}).
The smoothness condition
allows to put $qf_j(qx)=0$ directly in the set of equations defining $f_j(x)$
and obtain $N$-level dressing problem for zero potential.
It is known to lead to a potential describing
fixed time $N$-soliton solution of the KdV equation.

The formula (\ref{e20}) does not work at $q>1$. In fact,
this region of $q$ may be reached by analytical
continuation of $q<1$ solutions and permutations:
$$
f_1(x, q^{-1}; k_1, k_2,\dots, k_N)=iqf_1(qx/i, q; k_N, k_{N-1},\dots, k_1),
$$
$$
f_2(x,q^{-1})=if_N(x/i,q),\;
f_3(x,q^{-1})=if_{N-1}(x/i,q),\;
\dots, \; f_N(x,q^{-1})=if_2(x/i,q),
$$
where $k_j$-dependence in the last relations is the
same as in the first formula. In Ref.\cite{r9} the $N=1$ case was analyzed
at $q>1$ and the presence of singularities was demonstrated.
Analogous situation takes place in general -- analytic continuation
of smooth potentials to imaginary axis creates singularities. The latter
are moving under the scaling transformation and this breaks
needed isospectrality of $L_j$'s. Note that we do not $q$-deform Heisenberg
equations of motion, so that quantum mechanical time evolution coincides
with the standard one.  However, the time evolution of
the potentials $U(x)$ as infinite-number soliton solutions of the KdV equation
requires special consideration.

In order to elucidate the construction, we consider $N=2$
$q$-periodicity in more detail. Basic equations
\begin{eqnarray}
f_1^\prime(x) +f_2^\prime(x)+f_1^2(x)-f^2_2(x)=k_1, \nonumber \\
f_2^\prime(x) +qf_1^\prime(qx)+f_2^2(x)-q^2f^2_1(qx)=k_2,
\label{e21}
\end{eqnarray}
and operators $A^+=(p+if_1)(p+if_2)T_q$ and $H=a_1^+a_1^-$ generate the algebra
\begin{equation}
HA^+-q^2A^+H=(k_1+k_2)A^+,\qquad
A^-H-q^2 H A^-=(k_1+k_2)A^-,
\label{e23}
\end{equation}
\begin{equation}
A^-A^+-q^4A^+A^-=q^2(k_1(1+q^2)+2k_2 )H+k_2(k_1+k_2).
\label{e24}
\end{equation}
By adding to $H$ some constant  one may rewrite equations (\ref{e23})
in the form (\ref{e14}).
Relations (\ref{e23}), (\ref{e24}) at $q=1$ define conformal algebra $su(1,1)$
and at $q\neq 1$ they represent a particular ``quantization"
of this Lie algebra (see, e.g., Ref.\cite{r12}).

Let us find $f_{1,2}(x)$  as formal series near $x=0$.
Consider first the singular solutions. Permitted singularity
has a pole character:
$$f_1(x)={a\over x} +\sum_{i=1}^\infty b_i x^{2i-1},\qquad
f_2(x)=-{a\over x} +\sum_{i=1}^\infty c_i x^{2i-1},$$
\begin{equation}
b_i+c_i=\sum_{j=1}^{i-1} {c_jc_{i-j}-b_jb_{i-j}\over 2i-1+2a},\quad
q^{2i}b_i+c_i=\sum_{j=1}^{i-1} {q^{2i}b_jb_{i-j}-c_jc_{i-j}\over 2i-1-2a},
\label{e25a}
\end{equation}
$$b_1={1\over 1-q^2}({k_1\over 1+2a}-{k_2\over 1-2a}),\quad
c_1={1\over 1-q^2}({k_2\over 1-2a}-{q^2k_1\over 1+2a}),$$
where $a$ is an arbitrary parameter. In general, the series diverge at $q\to
1$,
proper choice of $a$, however, provides the truncated solution (\ref{e7}).
In the limit $q\to 0$, the function
$qf_1(qx)$ does not vanish, $qf_1(qx)\to a/x$. Substituting this into
(\ref{e21})
one gets two-level dressing problem for the potential $\propto 1/x^2$.
We guess that at $q<1$ there exist such $k_{1,2}$ that series converge
for arbitrarily large $x$. The condition $a(a+1)\geq 3/4$ guarantees that
normalizable wave functions and their first derivatives vanish at zero
\cite{r7a}. The spectrum of such system would arise from only one
geometric series (second is eliminated by boundary conditions).

For non-singular at zero solutions one has
$f_{1,2}=\sum_{i=0}^\infty b_i^{(1,2)} x^i,$
where $b_0^{(1,2)}$ are two arbitrary constants. Again, in general series
diverge at $q\to 1$. Particular choice of initial conditions gives
the solution which in this limit corresponds to (\ref{e7})
with a coordinate shift.
Depending on the values of $k_1$ and $k_2$ the limit $q\to 0$ recovers either
the smooth, two-level potential with bound state energies at $0$ and $k_1$,
or its analytically continued partner. Only first of these corresponds to
(\ref{e20}) at $N=2$, each series belonging to independent representation
of $su_q(1,1)$. Moreover, it is this solution that
reduces to the $q$-oscillator one
(\ref{e11})-(\ref{e14}) (with $q$ replaced by $q^{1/2}$)
after restrictions $f_2(x)=q^{1/2}f_1(q^{1/2}x),\; k_2=qk_1$.
At $q\to 1$ spectral series become equidistant which means that
potentials start to be unbounded at space infinities.
Because of the nice connection with ordinary conformal model \cite{r7a}, we
suggest to call the $N=2$ system as $q$-deformed conformal quantum mechanics.

For the $N=3$ system of equations:
\begin{eqnarray}
f_1^\prime(x) +f_2^\prime(x)+f_1^2(x)-f^2_2(x)=k_1, \nonumber \\
f_2^\prime(x) +f_3^\prime(x)+f_2^2(x)-f^2_3(x)=k_2, \nonumber \\
f_3^\prime(x) +qf_1^\prime(qx)+f_3^2(x)-q^2f^2_1(qx)=k_3,
\label{e26a}
\end{eqnarray}
one can exclude $f_{2,3}(x)$ and get
\begin{equation}
f_{2,3}(x)=-{1\over 2}f(x)\mp {1\over 2f(x)}(f^\prime(x)+k_2),
\label{e26b}
\end{equation}
\begin{equation}
f(x)={1+\sqrt{q}T_q\over 2}f_1(x)+{1-\sqrt{q}T_q\over 2}\int_0^x f_1^2(y)dy -
{\omega\over 2}(x-x_0),
\label{e26c}
\end{equation}
\begin{equation}
f^{\prime\prime}(x)={{f^\prime}^2(x)\over 2f(x)} +2f(x)(f_1^2(x)+
f_1^\prime(x)-\frac{1}{4}f^2(x)-f^\prime(x)-k_1-\frac{1}{2}k_2)
- {k_2^2\over 2f(x)},
\label{e26d}
\end{equation}
where $x_0$ is a constant of integration and $\omega = k_1+k_2+k_3$.
At $q=1$ one has $f_1(x)=f(x)
+\omega(x-x_0)/2$ and (\ref{e26d}) becomes the PIV equation (\ref{e9}).
So, the system (\ref{e26c}), (\ref{e26d}) describes a $q$-deformation
of the PIV transcendent \cite{r12a}.
In fact, all functions $f_i(x)$ satisfy one
combersome equation with different choices of the parameters $k_i$.
As a result, the relations (\ref{e26b}) give new solutions of the $q$-PIV
system in terms of a known one:
$f_2(x; k_1, k_2, k_3)=f_1(x; k_2, k_3, q^2k_1),\,
f_3(x; k_1, k_2, k_3)=f_1(x; k_3, q^2k_1, q^2k_2)$. Existence of these
nonlinear maps is a result of the hidden self-similarity \cite{adler}.

The notion of $q$-periodicity (\ref{e15}) and corresponding algebraic
relations (\ref{e17}), (\ref{e18}) constitute main results of
this paper. However,
above we just outlined properties of the self-similar potentials.
It is quite interesting to know what kind of potentials one gets as
a result of deformation of the finite-gap potentials, i.e. when for some
$j$ one has $k_j+k_{j+1}+\dots + k_{j+N-1}=0$.
There are other possibilities in addition to the mentioned ones.
For example, the coordinate $x$ and parameter $q$ were taken to be real
and nothing prevents
from the consideration of complex values as well.
An interesting situation is described when $q$ is a root of unity, $q^m=1$.
{}From the relations (\ref{e15}) one easily sees that now
\begin{equation}
f_{j+mN}(x)=q^mf_j(q^m x)=f_j(x),\qquad k_{j+mN}=q^{2m}k_j=k_j,
\label{e27a}
\end{equation}
which is a subcase of (\ref{e4}) because
$$\lambda_{j+mN}-\lambda_j  = (1-q^{2m})(k_j+k_{j+1}+\dots
+k_{j+N-1})/(1-q^2)=0. $$
At $N=1,\, q^3=1$ the simplest Lam\'e equation for the equianharmonic
Weierstrass function is arising \cite{root}.
Corresponding spectral problem is
known to be solvable \cite{per}. A charming property of the self-similar
systems in these cases is that operationally they are naturally characterized
not by the generators of the order $mN$ polynomial algebras but rather by
their $m$-th operator roots which are well defined and satisfy simpler
(although
unusual) commutation relations. Unfortunately, the general analytical
structure of the $q$-transcendents is not known  and this does
not allow to analyze analytical continuation of the solutions found
for the specific values of $q$. Also, there should exist some infinite-gap
potentials which are reduced to the self-similar ones in the limit of zero
widths of the gaps.

Different problems appear when the scaling operator $T_q$ (\ref{e12a}) in
the definition of $A^\pm$ is replaced
by the translation operator, $T_a f(x)=f(x+ a).$ Instead of
(\ref{e12}) we then have
\begin{equation}
f^\prime(x)+f^\prime(x+a)+f^2(x)-f^2(x+a)=k.
\label{e27}
\end{equation}
This equation may be considered also as a special limiting case of (\ref{e12}).
Since $q=1$, at $k\neq 0$ we have a realization of the ordinary Heisenberg-Weyl
algebra. The full effect of a non-zero parameter $a$
in (\ref{e27}) is not known to the author, but purely monotonic
solutions are seen to be forbidden. The particular solution at $k=0$
is given by
\begin{equation}
f(x)=-\,\frac{1}{2}\, {\wp'(x-x_0)-\wp'(a)\over \wp(x-x_0)-\wp(a)}\, ,
\label{e28}
\end{equation}
where $\wp(x)$ is an arbitrary doubly periodic Weierstrass function
(note that the values of corresponding periods do not enter the equation
(\ref{e27}), which exhibits a non-uniqueness of the solutions).

Higher dimensional generalizations of the presented construction are
not known to the author (except of the simple cases when variables
separate and the problem becomes effectively one-dimensional).
Several concluding remarks are in order.
First, the factorization method allows to
replace superpotentials $f_j(x)$ by hermitian matrix functions \cite{r13},
in which case right hand sides of Eqs. (\ref{e1}), (\ref{e13})
are proportional to unit matrices. Second, there may exist an interesting
interrelation between the described self-similar potentials and the wavelet
analysis \cite{chui} where affine transformations
generate orthonormal bases of the Hilbert space. Finally, the splitting
of the spectrum consisting of one geometric series (\ref{e14}) into the one
with $N$ terms (\ref{e20})
(at $q=1$ there is a splitting of the spectrum of a harmonic oscillator,
i.e. of the arithmetic progression \cite{r7})
illustrates the possibility for searching solutions of
the dressing chain (\ref{e1})
not by expansion in $j$ (or some function of $j$),
as it was done in Ref.\cite{r2}, but by representation $j=lN+i-1,\,
i=1,\dots, N,\, l\in Z$ and expansion in $l$.

\bigskip

The results of this paper were partially presented at the XIXth
International Colloquium on Group Theoretical Methods in Physics
(Salamanca, June 1992).
The author is indebted to J.-M.Lina, A.Shabat, and L.Vinet for valuable
discussions and helpful comments.
This research was supported by the NSERC of Canada.


\end{document}